# Superhyperfine interactions in $Ce^{3+}$ doped $LiYF_4$ crystal: ENDOR measurements

M.R. Gafurov [1], G.V Mamin [1], I.N. Kurkin [1], A.A. Rodionov [1], S.B. Orlinskii [1]

[1] Kazan Federal University, Kremlevskaya, 18, Kazan 420008, Russia

*E-mail: Marat.Gafurov@kpfu.ru

The first observation of the resolved Mims electron - nuclear double resonance (ENDOR) spectra from the nearby and remote nuclei of $^{19}F$ and $^{7}Li$ nuclei on impurity $Ce^{3+}$ ions in $LiYF_4$ crystal is reported. It shows that $LiYF_4$:$Ce^{3+}$ system can be exploited as a convenient matrix for performing spin manipulations and adjusting quantum computation protocols while ENDOR technique could be used for the investigation of electron-nuclear interaction with all the nuclei of the system and exploited for the electron - nuclear spin manipulations

## 1. Introduction

As a significant class of rare earth compounds, rare earth (RE) fluorides (REF) have been become anew a research focus in the material science due to their unique applications in optical communications, three-dimensional displays, solid-state laser, catalysis, solar cells, biochemical probes and medical diagnostics. They also has been regarded as the excellent host lattices for performing multicolor upconversion luminescence of the doped RE ions, due to their low phonon energy, high chemical stability and good optical transparency over a wide wavelength range. [1, 2].

Apart from being the cornerstone of modern laser technology, crystals and glasses doped with RE metals are known to have excellent properties for classical and quantum information storage and for performing quantum computation protocols [3 - 5]. In addition, $LiYF_4$ crystals, having a relatively simple crystalline structure and growing almost without defects, are unique model systems for studying the crystal field, electron-phonon, interionic, hyperfine interactions, and also isotopic effects [6]. To make good use of the excellent properties of REF, their morphology and microstructure should be carefully controlled. One of the powerful method of the RE doped crystals inspection is an electron paramagnetic/spin resonance (EPR/ESR). The interaction of paramagnetic ions in crystals with the magnetic moments of the ligand nuclei (superhyperfine interaction) usually leads to broadening of the EPR lines, their overlap and even the impossibility of observing the allowed superhyperfine structure (SHS) in the spectra [6]. Analysis of the SHS allows one to obtain important information on inter-ion interactions and distribution of electron spin density in crystals. For fluoride systems, SHS can vary in a huge range even in the same ligand environment depending on the paramagnetic ion: from 10 MHz for RE ions up to 300 MHz for $Pb^{3+}$ [7, 8]. The nuclear Zeeman splitting for $F^-$ in the X - band (about 9 GHz) range is also of about 10 MHz that complicates the decoding of the EPR spectra

To the best of our knowledge, up to now in $LiYF_4$: $Ce^{3+}$ systems the SHS with only the nearest fluorine environment (first coordination sphere) has been experimentally and theoretically investigated. Here we show that exploiting pulsed electron-nuclear double resonance (ENDOR) techniques one can study and use the superhyperfine interaction with the remote nuclei.

## 2. Experimental Part

Crystals of $LiYF_4$ doped with $\approx 0.05$ at % of $Ce^{3+}$ were grown in Kazan Federal University by the Bridgman-Stockbarger technique in argon atmosphere. EPR investigations were done by using helium flow cryostats by using X - band ESP 300 and Bruker Elexsys E580 spectrometers (T = 5-25 K, microwave frequency $\nu \approx 9.5$ GHz). We used two-pulse field swept electron spin echo (FS ESE) measurements for the detection of the primary Hahn–echo ($\pi/2 - \tau -\pi -$ echo), where $\tau$ is the interpulse delay time of 240 ns, with initial π/2 and π pulse lengths of 18 and 36 ns, respectively. For ENDOR experiments we used special double (for nuclei and electron) cavities and Mims pulse sequence (π/2 − τ − π/2 − T − π/2) with an additional radiofrequency (RF) pulse $π_{RF}$ = 18 μs inserted between the second and third microwave π/2 pulses. RF frequency in our setup could be swept in the range of 1 - 200 MHz. In the presented ENDOR results value of the magnetic field ***B*** was kept constant, corresponding to the orientation of ***B*** ⊥ *c* (*c* is the optical axis of crystal). Some basics of ENDOR are briefly described, for example, in our papers [9, 10].

## 3. Results and Discussion

Conventional EPR spectrum from $Ce^{3+}$ ions is shown in Fig. 1. Resonant magnetic fields in the orientations ***B*** ∥ *c* and ***B*** ⊥ *c* correspond to the *g* factors given in ref. [11]: $g_∥$ = 2.737; $g_⊥$ = 1.475. For $Ce^{3+}$ ions none of the stable isotopes ($^{136}Ce$, $^{138}Ce$, $^{140}Ce$, $^{142}Ce$) has a nuclear spin and, therefore, no hyperfine structure (HFS) exists.

The position of ENDOR spectrum and ENDOR splitting, $a_{ENDOR}$, can help not only to identify a type of nuclei coupled with the electron spins but also provide spatial relationships between them. Additionally, for $I ≥ 1$ an electric nuclear quadrupole coupling exists that can split or shift the ENDOR lines: the nuclear quadrupole interaction is sensitive to the electric field gradient at the site of the nucleus. While for the nearby nuclei a scalar coupling (also known as Fermy contact coupling or influence of the covalent bonding) should be taken into account, for the remote nuclei the pure electron - nuclei dipole - dipole interaction is often a good approximation. Indeed, earlier conventional ENDOR measurements for rare-earth ions in cubic and also in tetragonal sites in $CaF_2$ crystals have shown very little covalent bonding between the rare-earth ion (including $Ce^{3+}$ one) and fluorine nuclei even for the nearest environment [12, 13].

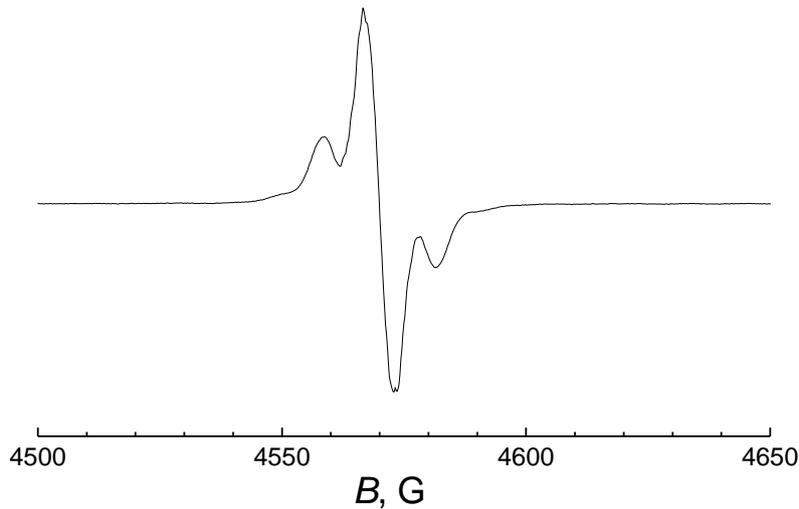

**Figure 1.** EPR line detected by conventional EPR technique with the SHS of $Ce^{3+}$ ions in $LiYF_4$. ***B***⊥*c*, T = 20 K, ν = 9.38 GHz.

All nuclei in $LiYF_4$ compound have sensible magnetic moment capable to create a perceptible additional magnetic field at the location of the RE impurity. In Tab. 1 some data for them which are important for the interpretation of ENDOR spectra are given.

ENDOR spectra for ***B*** ⊥ *c* presented in Fig. 2 are clearly due to the electron - nuclear interactions with $^7Li$ and $^{19}F$ isotopes. For other crystal orientations the ENDOR spectra shift correspondingly to the ***B*** value (data are not shown). The ability to observe narrow ENDOR lines at various RF frequencies from the Larmor ones testifies that $Ce^{3+}$ is an appropriate probe to sense remote nuclei.

The knowledge of the nuclear gyromagnetic ratios (Table 1) permits to define the values of *g* components obtained by the stationary method precisely: $g_{\parallel} = 2.7374$, $g_{\perp} = 1.4752$.

**Table 1**. Nuclear spin, natural abundance, relative gyromagnetic ratio and calculated Larmor frequency in $B = 4720$ G for the $LiYF_4$ nuclei

| Nuclei | Isotope | I | Natural Abundance, % | Relative to $^1H$ gyromagnetic ratio | Larmor frequency (B = 4720 G), MHz |
|---|---|---|---|---|---|
| $^3Li$ | 6 | 1 | 7.6 | 0.1471 | 2.96 |
|  | 7 | 3/2 | 92.4 | 0.3887 | 7.81 |
| $^9F$ | 19 | 1/2 | 100 | 0.9413 | 18.91 |
| $^{39}Y$ | 89 | 1/2 | 100 | 0.0492 | 0.99 |

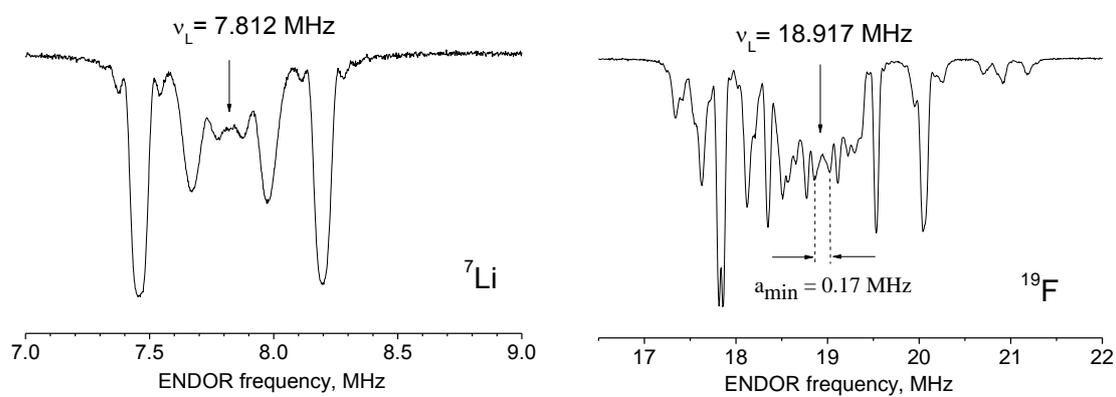

**Figure 2.** ENDOR spectra caused by interactions with $^7Li$ (left panel) and $^{19}F$ (right panel) isotopes in $LiYF_4$: $Ce^{3+}$ monocrystals for $\boldsymbol{B} \perp c$ orientation. B = 4721 G, T = 10 K, $\nu = 9.75$ GHz. Corresponding Larmor frequencies are marked by arrows. Dashed lines mark the minimal observed ENDOR splitting for $^{19}F$.

Technical limitations did not allow us to register low frequency (RF frequency of about and less than 1 MHz) ENDOR features which probably exist due to the presence of $^{39}Y$ nuclei. Interpretation of $^7Li$ ENDOR is complicated due to the influence of quadrupole interaction ($I = 3/2$, see Tab. 1) and is on the way.

Following the results and conclusions of papers [14, 15] for zinc oxide crystals, we suppose that the observed multitude of $^{19}F$ ENDOR lines in Fig. 2 indicates that we are dealing with a delocalized electron spin density of $Ce^{3+}$ ions. Exact estimation of the electron - nuclear distance to the first neighbours and to the most remote nuclei is a quite demanding task [16]. A rough estimation in the simplified ball-ball dipole-dipole approach gives the electron - nuclear distance to the most remote nuclei as $r \approx 0.66$ nm. Taking into account that in the $YF_8$ octahedra of $LiYF_4$ structure, the four first F neighbours are situated at 0.224 - 0.226 nm ($a_{ENDOR} \approx 4.2$ MHz), the distance to the four second neighbours is of about 0.229-0.232 nm [8], one can see that the electron spin density of $Ce^{3+}$ ions distributes well out of even the second coordination sphere. We hope that high frequency (microwave frequency of 94 GHz) experiments will shed light on the distribution of the electron density and will give the detailed information of electron-nuclear interactions in fluorides.

The rich, well-resolved, intensive ENDOR spectra show that $LiYF_4$:$Ce^{3+}$ system potentially can be used as an element (or model system) of the many qubit quantum computer. Indeed, it is known that electrons are natural candidates as physical qubits to be exploited for quantum computing and information processing (QC/QIP), and therefore magnetic resonance techniques, consisting of EPR and NMR are among the most appropriate techniques to be exploited for quantum computing. Pulsed EPR enables manipulation of electron spin and nuclear spin qubits in an equivalent manner and ENDOR can be used as the most useful spin manipulation technology in implementing QCs/QIPSs. The comprehensive review of the application of ENDOR techniques in quantum computations up to date is given in [17]. But to the best of our knowledge, up to now REF was not even considered as an appropriate system for the realization of many qubits operations.

## Acknowledgments

The work is financially supported by the Russian Science Foundation, Project # 17-72-20053.